\documentclass[12 pt,aps,a4paper]{article}
\usepackage{enumerate}
\usepackage{amssymb}
\usepackage{soul}
\usepackage{diagbox}
\usepackage{setspace}
\usepackage[english]{babel}
\usepackage{a4wide}
\usepackage{graphicx}
\usepackage{float}
\usepackage{amsthm}
\usepackage{graphics}
\usepackage{xcolor}
\usepackage{setspace}
\usepackage{subcaption}
\usepackage{mathtools}

\usepackage{afterpage}
\usepackage{indentfirst}
\usepackage{braket}
\usepackage{bm}
\usepackage{microtype}
\usepackage{wasysym}
\usepackage{array}
\usepackage{amsmath}
\usepackage{dsfont}

\begin{document}

\title{Hierarchy of quantum correlations in qubit-qutrit axially symmetric states}

\date{}

\maketitle

\begin{center}
\author{
Venkat Abhignan$^{*}$\\
Qdit Labs Pvt. Ltd., Bengaluru 560092, Karnataka, India \\
\and
R. Muthuganesan$^{\dagger}$\\
Department of Physics and Nanotechnology,\\
SRM Institute of Science and Technology, Kattankulathur, India
}
\end{center}

\begin{center}
$^{\dagger}$Corresponding author. E-mail: rajendramuthu@gmail.com \\
\end{center}

\begin{abstract}
We investigate quantum correlations in a hybrid qubit–qutrit system subject to both axial and planar single-ion anisotropies, dipolar spin–spin interactions, and Dzyaloshinskii–Moriya (DM) coupling. Using Negativity, Measurement-Induced Nonlocality (MIN), Uncertainty-Induced Nonlocality (UIN), and Bell nonlocality (as quantified by the CHSH inequality) as measures, we analyze the interplay between anisotropy parameters, magnetic fields, and temperature on the survival of quantum correlations. Our results demonstrate that Bell nonlocality and entanglement (Negativity) are highly sensitive to temperature and anisotropy, exhibiting sudden death under thermal noise, whereas MIN and UIN are significantly more robust. \textcolor{black}{ In particular, these discord-like and information-theoretic measures provide the largest baseline and persist even in parameter regions where entanglement vanishes, highlighting its suitability as a quantumness witness in realistic conditions. Notably, our Bell nonlocality study is tailored to the asymmetric qubit–qutrit setting by exploiting a recently developed qubit–qudit CHSH maximization framework. However, Bell nonlocality is confirmed to be the most fragile, surviving only in narrow parameter windows at low temperature. A key finding of this work is that we observe the fragility hierarchy
\[
\textcolor{black}{\text{Bell nonlocality} \subseteq \text{Negativity} \subseteq \text{UIN} (\text{MIN})}
\] in the qubit–qutrit setting.} These results provide deeper insight into the relative robustness of distinct quantum resources in anisotropic qubit–qutrit models, suggesting that quantum discord-like measures, such as MIN and UIN, may serve as more practical resources than entanglement for quantum information tasks in thermally active spin systems.
\end{abstract}

\noindent {\bf Keywords:}  Qubit-qutrit spin system; Quantum Correlations; Bell Nonlocality; Fragility Hierarchy; Thermal Noise.

                        
\maketitle


\section{\label{sec:level1}Introduction}

Quantum entanglement and, more generally, quantum correlations lie at the heart of quantum information science, enabling tasks such as secure communication and enhanced metrology that have no classical analog~\cite{Horodecki2009}.  While entanglement is a key resource (e.g.\ for teleportation and cryptography), it is now well understood that quantum correlations can exist even in separable mixed states~\cite{Ollivier2001}.  In fact, quantum entanglement is not only a special kind of nonclassical correlation, but separable states can also exhibit quantumness that can be exploited in information processing.  This has motivated the introduction of several measures beyond entanglement, such as quantum discord and its geometric variants, which remain nonzero at higher temperatures or noise levels where entanglement vanishes \cite{RevModPhys.84.1655}.   Furthermore, different versions of nonclassical correlation measures have been identified based on various notions, such as MIN \cite{Luo2011}, UIN \cite{Wu_2014}, measurement-induced disturbance (MID) \cite{Luo2008MID}, and local Fisher information–based measures \cite{Muthuganesan_2021,LQFI2018}. To effectively harness these measures for quantum information processing, a comprehensive understanding of their behavior in physical systems is crucial. In realistic physical systems, thermal fluctuations tend to suppress quantum coherence: for example, in mixed-spin chains, the thermal entanglement is found to decrease monotonically with temperature, eventually disappearing above a critical temperature that is largely independent of external fields~\cite{Wang2002,Wang2009}.  Nonetheless, more general quantum correlations can persist.  Remarkably, in some models, discord-like correlations even increase with temperature under a magnetic field (see e.g. studies on mixed-spin dimers and related models)~\cite{Naveena2022,Cencarikova2020}.  These observations suggest that different quantifiers capture distinct aspects of ``quantumness" in a thermal state, motivating a comprehensive comparison of such measures within a single model.

The practical relevance of this distinction is broad: entanglement underpins protocols such as teleportation and entanglement-enhanced metrology via spin-squeezing~\cite{Horodecki2009,Kitagawa1993}. At the same time, more robust discord-type or information-theoretic correlations (e.g.\ LQU~\cite{Wu_2014} or LQFI \cite{LQFI2018}) can remain useful when entanglement is destroyed by thermal noise.  Recent work by Yurischev \textit{et al.} derived compact closed forms for LQU and LQFI in axially symmetric qubit-qutrit states and reported cascades of sudden changes in these measures as temperature or interaction parameters vary, highlighting qualitatively different thermal responses of these distinct quantifiers~\cite{Yurischev2025}. {\color{black} Haddadi \textit{et al.} further extended these results to general qubit–qudit axially symmetric systems and reported nontrivial thermal behavior of LQU and LQFI, including correlation enhancement in specific temperature regimes \cite{Haddadi2025}. More generally, Yurischev \textit{et al.} have investigated spin-$(j,1/2)$ models and shown that entanglement thresholds shrink to zero as $j \to \infty$. In contrast, LQU and LQFI remain robust even under small symmetry-breaking perturbations \cite{qxtm-gysy}. Closed-form expressions for LQFI and LQU are also known for general two-qubit $X$ states \cite{YURISCHEV2023}, further illustrating the availability of analytical tools for these measures. In contrast to these studies, the present work focuses on MIN, UIN, entanglement, and Bell nonlocality, which probe fundamentally different aspects of quantum correlations.} MIN and UIN provide complementary geometric and information-theoretic perspectives on such surviving correlations~\cite{Luo2011,Wigner1963,Chaouki2023}.  For concrete mixed-spin materials and heterostructures, DM and related spin–orbit terms play an important role in setting the microscopic Hamiltonian and can strongly influence the correlation landscape~\cite{Dzyaloshinskii1958,Moriya1960,Sinova2015,Zutic2004,Wolf2001}.  Taken together, these theoretical and model studies motivate a unified investigation of Negativity (entanglement), MIN, UIN, and Bell nonlocality in a realistic mixed spin-($1/2,1$) Hamiltonian. Hence, it is necessary to chart which resources remain viable under thermal noise and which rapidly vanish.

Recent theoretical work has begun to explore these diverse correlation measures in explicit qubit–qutrit models. For example, closed-form linear-entropy discord and MID were derived for a qubit–qutrit spin chain at finite temperature and in an external magnetic field, showing that discord and MID remain nonzero well beyond the point where entanglement (logarithmic Negativity) vanishes and can even signal critical points in the system \cite{Benabdallah2020}. In a related study, qubit–qutrit pairs were studied under global dephasing noise and observed that discord and MID can “freeze” while entanglement suffers sudden death \cite{Benabdallah2021}. Further studies extended these analyses to general collective dephasing of qubit–qutrit states, identifying conditions under which both entanglement and LQU persist in the asymptotic steady state \cite{Ali2020}. Spin–orbit effects have also been considered where qubit–qutrit have been analysed in an XY spin chain with Dzyaloshinskii–Moriya coupling, finding that the DM interaction accelerates the decay of quantum correlations and that discord (and MID) is the most sensitive indicator of quantum phase transitions compared to entanglement \cite{Yang2014}. More recently, a XXZ qubit–qutrit model with intrinsic decoherence was studied, along with DM coupling, and inhomogeneous fields, comparing entanglement, LQU, and UIN. They found that UIN remains nonzero and detects quantum correlations even when entanglement and LQU have decayed to zero \cite{Chaouki2023}. {\color{black}Qubit-qutrit systems have also been analyzed in open-system settings, studying coherence dynamics in a dissipative environment~\cite{Manan_2025}.} Together, these results reinforce that in realistic qubit–qutrit systems, non-entanglement correlations (discord, MID, UIN, etc.) often survive thermal and decohering effects much longer than Negativity, motivating a detailed comparison of these measures in our anisotropic mixed-spin model.

As mentioned, we quantify the nonclassical correlations of the thermal state, using four complementary measures.  First, we utilize the Negativity, a computable entanglement measure defined via the Peres--Horodecki positive partial transpose (PPT) criterion~\cite{Peres1996,Horodecki1996}.  Negativity is given by the sum of the negative eigenvalues of the partially-transposed density matrix, and $\mathcal{N}>0$ is a necessary and sufficient signature of entanglement for $2\times3$ systems. {\color{black} This is a special feature of only $2\times3$ systems, where negativity fully characterizes entanglement. In higher-dimensional systems, however, the PPT criterion is no longer sufficient. Some entangled states remain PPT and are not detected by Negativity (so‐called bound entangled states \cite{PhysRevLett.80.5239}), which has been discussed in contexts such as spin-1 Heisenberg dimer models~\cite{Benabdallah2025qip}.} As a bipartite entanglement measure, Negativity captures the purely inseparable resources available in the state.  Second, we consider MIN, which was introduced by Luo and Fu as a geometric measure of quantum correlation~\cite{Luo2011}.  MIN quantifies the maximal global disturbance to the state induced by locally non-disturbing projective measurements on the qubit.  In other words, MIN measures how much the overall state changes under the optimal local measurement that preserves the qubit’s reduced density matrix.  Unlike entanglement, MIN can be nonzero for separable states and thus captures a broader class of nonclassical correlations.  Third, we consider the related concept of UIN.  UIN is defined via the Wigner--Yanase skew information~\cite{Wigner1963} and quantifies the maximum quantum uncertainty in local observables that commute with the reduced state.  It can be viewed as an improved version of MIN (resolving certain mathematical issues) that still admits a closed-form expression for $2\times3$ systems.  Both MIN and UIN remain invariant under local unitary transformations and typically decay more slowly with temperature than entanglement. Previously, we showed that UIN and LQU decay more slowly than entanglement in two-qubit spin squeezing models subject to intrinsic decoherence~\cite{Abhignan2021}. {\color{black}Finally, we compute the Bell nonlocality by evaluating the maximal CHSH Bell-inequality violation for the qubit--qutrit state using the general qubit–qudit CHSH framework of Bernal et al \cite{g3mw-6rl1}.}  A value of the CHSH parameter $>2$ certifies that the state cannot be described by any local hidden-variable model, representing the strongest form of bipartite nonclassicality.  \textcolor{black}{In practice, only very pure entangled states violate Bell inequalities, making Bell nonlocality the most fragile correlation measure.}

By analyzing all four of these measures, we can map out the hierarchy of quantum correlations in the mixed-spin system.  Previous studies of similar systems have found that as temperature increases or noise accumulates, Bell nonlocality is lost first, entanglement survives to higher temperatures, and even more general correlations (discord-type measures) can persist the longest~\cite{Horodecki2009,Ollivier2001}.  We indeed observe such an ordering in our model: MIN and UIN detect nonclassicality in parameter regimes where Negativity is already zero, and Bell violation is below the classical bound.  Our goal is to understand quantitatively how each interaction parameter in the Hamiltonian (exchange couplings, anisotropies, fields, etc.) influences this hierarchy.  This is important from both fundamental and practical perspectives: different quantum information tasks may rely on different resources, and thermal robustness is crucial for realistic implementations.  For example, entanglement-based tasks (e.g.\ teleportation) require Negativity $>0$, whereas specific metrological or communication protocols might harness any nonzero MIN/UIN (quantum discord \cite{Pirandola2014}).  By providing a unified study of entanglement, MIN, UIN, and Bell nonlocality in a physically relevant model, we aim to clarify which forms of quantumness remain viable under thermal noise and which are most resilient in mixed-spin materials.

The paper is organized as follows.  In Sec.~2, we present the axially-symmetric Hamiltonian for the qubit--qutrit model and derive its thermal Gibbs state analytically based on Ref.\cite{Yurischev2025}.  In Sec.~3, we review the definitions of Negativity, MIN, UIN, and Bell nonlocality and discuss their computation for a $2\times3$ thermal state.  In Sec.~4, we present numerical results illustrating the temperature and parameter dependence of these measures, highlighting their distinct behaviors.  We find that, consistent with general expectations, Bell nonlocality is the most fragile and is present only at very low $T$, whereas MIN and UIN remain significant even when entanglement is gone.  Our results demonstrate the value of these measures for capturing different facets of quantum correlations in a realistic spin model, and suggest that robust tasks could exploit the longer-lived MIN/UIN correlations rather than relying solely on entanglement.  Comprehensive parameter scans allow us to identify the conditions under which each form of quantumness is maximized or suppressed.  Finally, Sec.~5 summarizes our findings and discusses potential implications for quantum information applications in mixed-spin systems.  

\section{The Model and Thermalization}

{\color{black}We study the thermal quantum correlations in a bipartite system made up of a spin-1/2 particle (qubit) coupled to a spin-1 particle (qutrit). We primarily consider an axially-symmetric qubit–qutrit Hamiltonian introduced in Ref. \cite{Yurischev2025}, which includes longitudinal fields and different couplings.}

The qubit operators are $s_i=\sigma_i/2$, where $\sigma_i$ are the Pauli matrices, while the qutrit (spin-1) operators are
\begin{equation}
S_x = \frac{1}{\sqrt{2}}\begin{pmatrix}0 & 1 & 0\\ 1 & 0 & 1\\ 0 & 1 & 0\end{pmatrix},
\quad
S_y = \frac{1}{\sqrt{2}}\begin{pmatrix}0 & -i & 0\\ i & 0 & -i\\ 0 & i & 0\end{pmatrix},
\quad
S_z = \begin{pmatrix}1 & 0 & 0\\ 0 & 0 & 0\\ 0 & 0 & -1\end{pmatrix}.
\end{equation} 
Defining the total $z-$component of spin as
\[
\mathcal{S}_z = s_z \otimes I_3 + I_2 \otimes S_z,
\]
with $I_n$ denoting the $n\times n$ identity matrix, we require that the Hamiltonian commute with $\mathcal{S}_z$, ensuring invariance under rotations about the $z$-axis:
\[
[\mathcal{H},\mathcal{S}_z] = 0.
\]
Subject to this symmetry, the most general Hamiltonian with real coupling constants is
\begin{equation}\label{eq:Hamiltonian}
\begin{split}
\mathcal{H} &= B_1\, s_z + B_2\, S_z 
 + J\,(s_x S_x + s_y S_y) + J_z\, s_z S_z 
 + K\, S_z^2 + K_1\,(S_x^2 + S_y^2) + K_2\, s_z S_z^2 \\
&\quad + D_z\,(s_x S_y - s_y S_x)
  + \Gamma\bigl[s_x(S_x S_z + S_z S_x) + s_y(S_y S_z + S_z S_y)\bigr] \\
&\quad + \Lambda\bigl[s_x(S_y S_z + S_z S_y) - s_y(S_x S_z + S_z S_x)\bigr].
\end{split}
\end{equation}
The parameters in the Hamiltonian represent various physically significant interactions in mixed-spin systems. The coefficients $B_1$ and $B_2$ correspond to the longitudinal magnetic fields acting on the qubit and qutrit, respectively, which control the system's magnetization \cite{Naveena2022,Hagiwara1999,Cencarikova2020}. The terms $J$ and $J_z$ denote the transverse and longitudinal components of the XXZ Heisenberg exchange interaction between the qubit and qutrit spins \cite{Naveena2022,Wang2009}. The constants $K$ and $K_1$ represent uniaxial and planar single-ion anisotropies affecting the alignment of the spin-1 particle. The biquadratic anisotropy term $K_2$ accounts for two-ion interactions studied in the context of spin-chain materials. The parameter $D_z$ introduces the DM interaction along the $z$-axis, which is essential for describing chiral magnetism and spin Hall effects \cite{Dzyaloshinskii1958,Moriya1960}. The higher-order parameters $\Gamma$ and $\Lambda$ describe three-spin couplings that arise in systems with strong spin--orbit interaction, leading to more complex spin dynamics. Collectively, these parameters enable a comprehensive description of realistic qubit--qutrit systems and their thermal quantum correlations.

In the standard computational basis, the Hamiltonian of the axially symmetric (AS) system, denoted by $\mathcal{H}$, takes the form \cite{Yurischev2025}
\begin{equation}
\mathcal{H} = 
\begin{pmatrix} E_1 & & & & & \\
 & h_1 & 0 & g_1 & 0 & \\
 & 0 & h_2 & 0 & g_2 & \\
 & g_1^* & 0 & h_3 & 0 & \\
 & 0 & g_2^* & 0 & h_4 & \\
 & & & & & E_6 \end{pmatrix},
\end{equation}
where the diagonal elements  are $ h_{1,4} = \pm \tfrac{B_1}{2} + 2K_1$, $h_{2,3} = \pm \tfrac{B_1}{2} \mp B_2 - \tfrac{J_z}{2} + K + K_1 \pm \tfrac{K_2}{2}$ and the off-diagonal elements are given by $g_{1,2}=\left(J \pm \Gamma + i(D_z \pm \Lambda)\right)/\sqrt{2}$.

Upon diagonalization, the eigenvalues of the Hamiltonian are obtained as
\begin{align}
E_{1,6} &= \frac{J_z}{2} + K + K_1 \pm \Bigl(\tfrac{B_1}{2} + B_2 + \tfrac{K_2}{2}\Bigr), \\
E_{2,3} &= \tfrac12(h_1+h_3 \pm R_1),\quad
E_{4,5} = \tfrac12(h_2+h_4 \pm R_2),
\end{align}
where the quantities $R_1$ and $R_2$ are defined as 
\begin{align}
R_1 = \sqrt{(h_1-h_3)^2 + 4|g_1|^2},\quad
R_2 = \sqrt{(h_2-h_4)^2 + 4|g_2|^2}.
\end{align}
Finally, the trace of the Hamiltonian is given by
\begin{align}
    \text{Tr}\mathcal{H} = 4(K+2K_1)
\end{align}
which remains independent of the magnetic field and spin–spin interaction parameters, reflecting the preserved axial symmetry of the model.

In thermal equilibrium at temperature $T$, the system is described by the Gibbs state
\begin{equation}
\rho = \frac{1}{Z}e^{-\mathcal{H}/T},\qquad Z = \mathrm{Tr}(e^{-\mathcal{H}/T}).
\end{equation}
The density matrix is of the form \cite{Yurischev2025} \begin{equation}
   \label{eq:rho}
   \rho=
	 \left(
      \begin{array}{cllccc}
      p_1&\ &\ &\ &\ \\
      \ &a&0&u&0&\ \\
			\ &0&b&0&v&\ \\
      \ &u^*&0&c&0&\ \\
      \ &0&v^*&0&d&\ \\
			\ &\ &\ &\ &\ &p_6
      \end{array}
   \right),
\end{equation}
 and analytic expressions for nonzero elements in the AS basis are:
\begin{align}
 p_1 &= \frac{1}{Z}e^{-E_1/T},\quad  p_6 = \frac{1}{Z}e^{-E_6/T}, \\
 a &= \frac{1}{Z}\Bigl(\cosh\tfrac{R_1}{2T} + \tfrac{h_3-h_1}{R_1}\sinh\tfrac{R_1}{2T}\Bigr)e^{-(h_1+h_3)/2T},\nonumber\\
 b &= \frac{1}{Z}\Bigl(\cosh\tfrac{R_2}{2T} + \tfrac{h_4-h_2}{R_2}\sinh\tfrac{R_2}{2T}\Bigr)e^{-(h_2+h_4)/2T},\nonumber\\
 c &= \frac{1}{Z}\Bigl(\cosh\tfrac{R_1}{2T} + \tfrac{h_1-h_3}{R_1}\sinh\tfrac{R_1}{2T}\Bigr)e^{-(h_1+h_3)/2T},\nonumber\\
 d &= \frac{1}{Z}\Bigl(\cosh\tfrac{R_2}{2T} + \tfrac{h_2-h_4}{R_2}\sinh\tfrac{R_2}{2T}\Bigr)e^{-(h_2+h_4)/2T},\nonumber\\
 u &= -\frac{2g_1}{Z R_1}\sinh\tfrac{R_1}{2T}e^{-(h_1+h_3)/2T},\quad
 v = -\frac{2g_2}{Z R_2}\sinh\tfrac{R_2}{2T}e^{-(h_2+h_4)/2T}.\nonumber
\end{align}
The partition function can be written explicitly as:
\[
Z = 2\Bigl[\cosh\tfrac{B_1+2B_2+K_2}{2T}e^{-(J_z+2K+2K_1)/2T} + \cosh\tfrac{R_1}{2T}e^{-(h_1+h_3)/2T} + \cosh\tfrac{R_2}{2T}e^{-(h_2+h_4)/2T}\Bigr].
\]

\section{Quantum Correlation measures} \label{thermodynamics}
This section reviews the quantumness measures addressed in this manuscript. To analyze the nonlocal characteristics of the axially symmetric state governed by the Hamiltonian in Eq. (\ref{eq:Hamiltonian}), we consider four measures: entanglement (quantified by Negativity), MIN, UIN, and Bell nonlocality. Let $\rho$ be the bipartite density matrix describing the thermal state of auxiliary symmetric state Eq. (\ref{eq:Hamiltonian}), defined on the separable Hilbert space $\mathcal{H}^a\otimes \mathcal{H}^b$. Here $\mathcal{H}^{a(b)}$ denotes the Hilbert space of the subsystem $\rho_{a(b)}=\text{Tr}_{b(a)}\rho$. 
\subsection{Negativity}
Entanglement is one of the most peculiar phenomena exhibited by composite quantum systems. Negativity is a widely used measure of bipartite entanglement, particularly suitable for higher-dimensional systems. It is based on the Peres–Horodecki criterion (PPT criterion).
 For a given bipartite state $\rho$, the partial transpose with respect to subsystem $a$ is defined as 
\begin{align}
    \langle ij|\rho^{T_a} | kl \rangle =  \langle kj | \rho | il \rangle \nonumber
\end{align}
For a separable state $\rho$, the partial transpose remains positive semidefinite. However, if  $\rho_{a}$ has negative eigenvalues, the state is entangled. The negativity quantifies the degree of this violation and is defined as
\begin{align}
    \mathcal{N}(\rho)=\sum_{i=1}^6\frac{1}{2}(|\lambda_i|-\lambda_i)
    \label{negativity}
\end{align}
where $\lambda_i$ are the eigenvalues of the partially transposed density matrix $\rho_{a}$. Negativity thus provides a simple and effective way to quantify the degree of entanglement present in a quantum state.
\subsection{Measurement-Induced Nonlocality}
The second measure of interest is MIN, which captures quantum correlations in a bipartite system beyond entanglement. Originally introduced by S. Luo and S. Fu, from a geometric perspective, MIN is defined as 
\begin{align}
   \text{ MIN} =~^\text{max}_{\Pi^a} || \rho-\Pi^a(\rho)||^2
\end{align}
where the post-measurement state is given as 
\begin{align}
  \Pi^a(\rho)=\sum_k(\Pi^a_k\otimes\mathds{1}^b) \rho (\Pi^a_k\otimes\mathds{1}^b)   \nonumber
\end{align}
with $\Pi^a_k$ being the eigenprojectors constructed from the eigenvectors of subsystem $a$. When computing MIN, the reduced states of the local subsystems remain unchanged. This invariance of the marginal states under local eigenprojective measurements makes MIN a more robust and secure resource for quantum information processing tasks compared to entanglement.  An arbitrary qubit-qutrit state in the Bloch-Fano representation can be written as 
\begin{equation}
\rho=\frac{1}{2}\left[\mathds{1}_2 \otimes \mathds{1}_3+\sum_{i=1}^3 x_i (\sigma_i\otimes \mathds{1}_3)+\sum_{j=1}^3 y_j (\mathds{1}_2 \otimes \Lambda_j)+\sum_{i,j} t_{ij} \sigma_i \otimes \Lambda_j\right],  \nonumber
\label{Equations} 
\end{equation}
where $x_i=\text{Tr}(\rho(\sigma_i\otimes \mathds{1}_3))$, $y_j=\text{Tr}(\rho(\mathds{1}_2 \otimes \Lambda_j))$ are the components of Bloch vector and $t_{ij}=\text{Tr}(\rho(\sigma_i \otimes \Lambda_j))$ represent real matrix elements of correlation matrix $T$. Here, $\sigma_i$ and $\Lambda_j$ are the spin-1/2 and spin-1 operators.

The closed formula of MIN for the qubit-qutrit system is 
\begin{equation}
\text{MIN} =
\begin{cases}
\text{Tr}(TT^t)- \frac{1}{|\text{x}|^2}\text{x}^tTT^t\text{x}~~~~
 \text{if}~~~~~~~~ \text{x} \neq  0, \\
\text{Tr}(TT^t)-\tau_{\text{min}} ~~ ~~~~~~~~\text{otherwise}.
\end{cases} 
\label{HSMIN}
\end{equation}
where $\tau_{\text{min}}$ is the least eigenvalue of matrix $TT^t$  and the matrix elements of the correlation matrix $T$ are defined as $t_{ij}=\text{Tr}[\rho(\sigma_i\otimes \Lambda_j)]$. 

\subsection{Uncertainty-Induced Nonlocality} \label{landlimit}

Similar to MIN, the UIN captures the global effect of local quantum measurements that leave the local state invariant but induce a maximal global disturbance on the composite system.  Formally, UIN for a bipartite state $\rho$ on $\mathcal{H}^a\otimes \mathcal{H}^b$ is defined using the skew information associated with local observables. The skew information of a state $\rho$ with respect to an observable 
$K$ is given by
\begin{equation}
    I(\rho, K) = -\frac{1}{2}\,\mathrm{Tr}\!\left([\sqrt{\rho}, K]^2\right),
\end{equation}
which quantifies the amount of quantum uncertainty of $K$ in the state $\rho$.

The UIN of $\rho$ with respect to local measurements on subsystem $a$ is then defined as
\begin{equation}
    \mathcal{U}(\rho) = \max_{K_a}\, I\!\left(\rho, K_a \otimes I_b\right),
\end{equation}
where the maximization runs over all local observables $K_a$ on $\mathcal{H}^a$ that commute with the reduced density matrix $\rho_a$.

For a general bipartite state $\rho$, UIN admits a closed expression in terms of the correlation matrix. The UIN is given by
\begin{equation}
    \mathcal{U}(\rho) = 1 - \lambda_{\min}(W W^T),
\end{equation}
where $W_{mn}=\text{Tr}(\sqrt{\rho}(\sigma_m\!\otimes\!\Lambda_n)$ is the correlation matrix of $\rho$ and $\lambda_{\min}(WW^T)$ denotes the smallest eigenvalue of $WW^t$. This expression provides an operationally accessible quantification of UIN, directly linked to the strength of quantum correlations present in the state.


\subsection{Bell Nonlocality} \label{bell_nonlocality}
Bell nonlocality provides an operational criterion to reveal correlations in a bipartite quantum system that cannot be explained by any local hidden-variable model \cite{Clauser1969,Pironio2014}. The CHSH Bell operator for a qubit–qutrit system is
\begin{equation}
\mathcal{B} \;=\; A_0\otimes(B_0 + B_1)\;+\;A_1\otimes(B_0 - B_1),
\end{equation}
where $A_{0,1}$ are Alice's qubit observables (eigenvalues$\pm1$) and $B_{0,1}$ are Bob's dichotomic qutrit observables (Hermitian involutions with eigenvalues$\pm1$, possibly degenerate). The CHSH expectation is $\langle\mathcal{B}\rangle=\mathrm{Tr}(\rho\mathcal{B})$, and the maximal CHSH value is $\mathcal{B}_{\max}(\rho)=\max_{A_{0,1},B_{0,1}}\langle\mathcal{B}\rangle$. The condition for Bell-nonlocal violation is
\[
\mathcal{B}_{\max}(\rho)\;>\;2,
\]
and the standard qubit–qubit maximization reduces to the Horodecki criterion \cite{Horodecki1995}, while the qubit–qudit maximization used here follows Bernal \emph{et al.} \cite{g3mw-6rl1}.

The qubit-qutrit state $\rho$ is represented in the operator–Schmidt form
\begin{equation}
\rho=\tfrac{1}{2}\Big(\mathbb{I}_2\otimes\beta_0+\sum_{i=1}^3\sigma_i\otimes\beta_i\Big),
\end{equation}
with $\beta_i=\text{Tr}_A[\rho(\sigma_i\otimes\mathbb{I}_d)]$ ($d=3$ for a qutrit). Furthermore, the full measurement optimization is reduced to a single optimization over rotations $R\in SO(3)$. Defining the rotated beta matrices by
\[
(R\boldsymbol{\beta})_a=\sum_{i=1}^3 R_{ai}\,\beta_i\qquad(a=1,2,3),
\]
and using the trace norm $\|M\|_1=\text{Tr}\sqrt{M^\dagger M}$ (for Hermitian $M$ equal to the sum of absolute eigenvalues), the maximal CHSH value can be written as
\begin{equation}
\mathcal{B}_{\max}(\rho)
\;=\;2\max_{R\in SO(3)}\sqrt{\;\big\|(R\boldsymbol{\beta})_1\big\|_1^{2}+\big\|(R\boldsymbol{\beta})_2\big\|_1^{2}\;}.
\label{eq:Bmax_final}
\end{equation}
Equivalently, if $\{\lambda^{(a)}_k(R)\}_{k=1}^{d}$ are the eigenvalues of $(R\boldsymbol{\beta})_a$, then
\[
\mathcal{B}_{\max}(\rho)=2\max_{R\in SO(3)}\Bigg\{\Bigg(\Big(\sum_{k=1}^d\big|\lambda^{(1)}_k(R)\big|\Big)^2+\Big(\sum_{k=1}^d\big|\lambda^{(2)}_k(R)\big|\Big)^2\Bigg)^{1/2}\Bigg\}.
\]

This formulation is both operational and numerically efficient, requiring only a three-parameter search to compute $\mathcal{B}_{\max}$. Once the optimal rotation $R^\star$ is found, the optimal Bob observables are obtained by diagonalizing $(R^\star\boldsymbol{\beta})_{1,2}$ and aligning their eigenvalue signs (trace-norm alignment); Alice's optimal observables follow the usual CHSH alignment with the vectors induced by these Bob operators. 

\section{Results and Discussions}
{\color{black}In what follows, we present a comparative study of quantum correlation measures for qubit–qutrit axially symmetric states. Yurischev et al. focused exclusively on the LQU and the LQFI for this hybrid qubit–qutrit model \cite{Yurischev2025}. Going beyond their analysis, we extend the investigation to include entanglement quantified by negativity, two discord-like measures MIN and UIN as well as Bell nonlocality, characterized via a generalized CHSH bound \cite{g3mw-6rl1}. We adopt the same parameter regime as in Ref. \cite{Yurischev2025} to enable a direct comparison and to highlight the correspondence between the behavior of our considered measures and that of LQU and LQFI.}

\begin{figure}[!h]
\centering
\begin{subfigure}[b]{0.45\textwidth}
    \centering
    \includegraphics[width=\textwidth, height=130px]{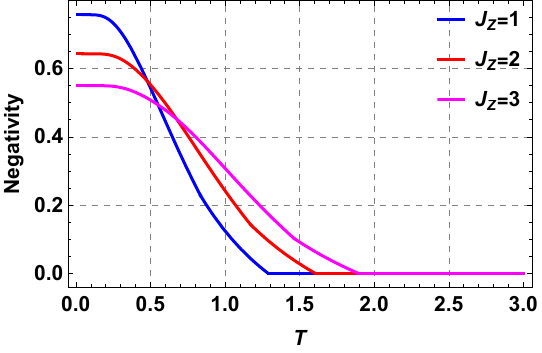}
    \caption{Negativity}
    \label{Fig1a}
\end{subfigure}
\hfill
\begin{subfigure}[b]{0.45\textwidth}
    \centering
    \includegraphics[width=\textwidth, height=130px]{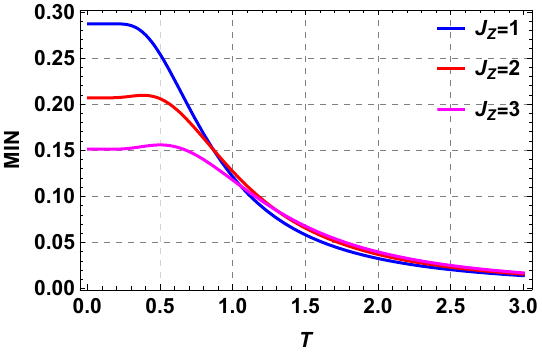}
    \caption{MIN}
    \label{Fig1b}
\end{subfigure}

\vspace{0.3cm}

\begin{subfigure}[b]{0.45\textwidth}
    \centering
    \includegraphics[width=\textwidth, height=130px]{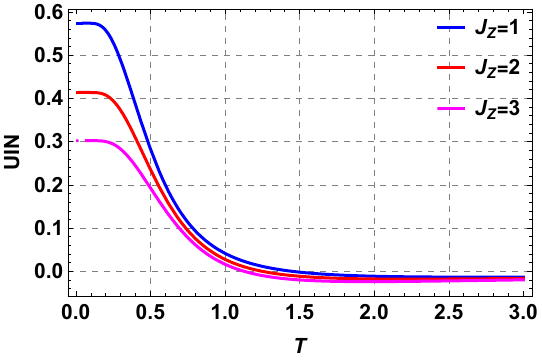}
    \caption{UIN}
    \label{Fig1c}
\end{subfigure}
\hfill
\begin{subfigure}[b]{0.45\textwidth}
    \centering
    \includegraphics[width=\textwidth, height=130px]{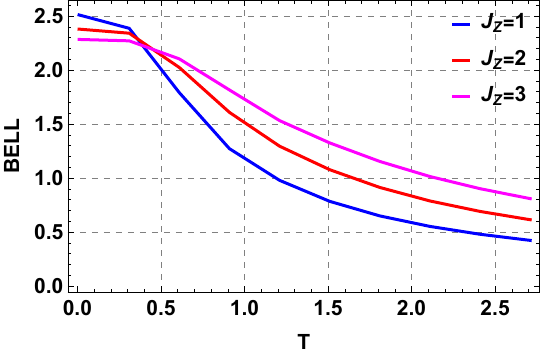}
    \caption{Bell (CHSH)}
    \label{Fig1d}
\end{subfigure}
\caption{Thermal quantum correlations for $B_1 = 0.3, B_2 = -0.7, J = 0, K = 0.2, K_1 = -0.1, K_2 = 0.22, Dz = 0.32, \Gamma = -0.87, \Lambda = 0.31.$ Panels show: (a) Negativity, (b) Measurement-Induced Nonlocality (MIN), (c) Uncertainty-Induced Nonlocality (UIN), and (d) Bell (CHSH) parameter.} 
\label{Fig1}
\end{figure}

In Fig. \ref{Fig1}, we initially examine the behaviors of entanglement quantified by negativity, MIN, UIN and Bell nonlocality as functions of temperature for the fixed parameters $B_1 = 0.3,~ B_2 = -0.7,~ J = 0,~ K = 0.2, ~K_1 = -0.1, ~K_2 = 0.22,~ Dz = 0.32,~ \Gamma = -0.87, ~\Lambda = 0.31.$ We vary the longitudinal exchange interaction $J_z = 1, ~2 ~\text{and}~~ 3$.

At $T=0$, the state is maximally entangled, yielding the highest Negativity. As $T$ increases, thermally induced mixing reduces the off‑diagonal coherences responsible for entanglement, causing a smooth, monotonic decay of Negativity. At finite temperature, “entanglement‑death” is reached when Negativity falls to zero. MIN also decreases with temperature, but it persists beyond entanglement‑death, reflecting that certain nonclassical correlations survive even in separable states. Its decay is gentler than that of Negativity, vanishing significantly only at higher temperatures. Defined via the Wigner–Yanase skew information, UIN begins at a larger value than MIN at $T=0$. Its thermal decay parallels that of MIN and Negativity, but does not exhibit crossover behavior between $ J_z=1,2,3$. Specifically, $J_z=1$ has the highest correlations at $T=0$ but decays first. {\color{black}For the same set of parameters, this $J_z=1$ decay mirrors the behavior of LQU and LQFI in Ref. \cite{Yurischev2025}, which likewise decrease without any sudden jumps as $T$ rises. A key difference is that Negativity and Bell–CHSH reach zero and cross the classical limit 2, respectively, at lower temperature, whereas LQU/LQFI (discord-like measures) approach zero more gradually.} The CHSH parameter attains its maximum only in the very low‑temperature regime. Even moderate thermal fluctuations reduce the Bell–CHSH value below the classical bound of 2, such that Bell nonlocality is lost. This hierarchy of fragility — $\text{Bell nonlocality} \subseteq \text{Negativity} \subseteq \text{UIN} (\text{MIN})$ — highlights the increasing robustness of more general nonclassical correlations against thermal noise. Furthermore, the hierarchy suggests that the UIN (MIN) is a better resource against thermal fluctuations in a thermally activated system. 

\begin{figure}[!h]
\centering
\begin{subfigure}[b]{0.45\textwidth}
    \centering
    \includegraphics[width=\textwidth, height=130px]{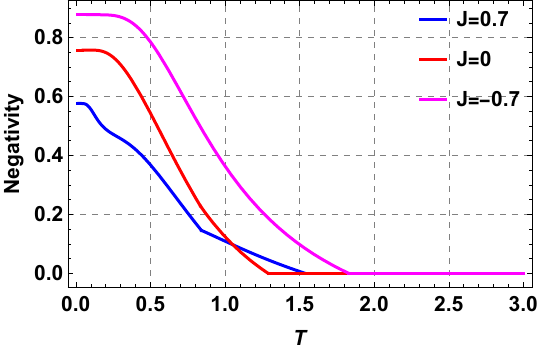}
    \caption{Negativity}
    \label{Fig1a}
\end{subfigure}
\hfill
\begin{subfigure}[b]{0.45\textwidth}
    \centering
    \includegraphics[width=\textwidth, height=130px]{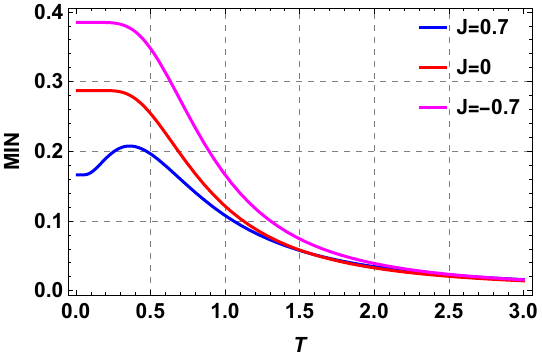}
    \caption{MIN}
    \label{Fig1b}
\end{subfigure}

\vspace{0.3cm}

\begin{subfigure}[b]{0.45\textwidth}
    \centering
    \includegraphics[width=\textwidth, height=130px]{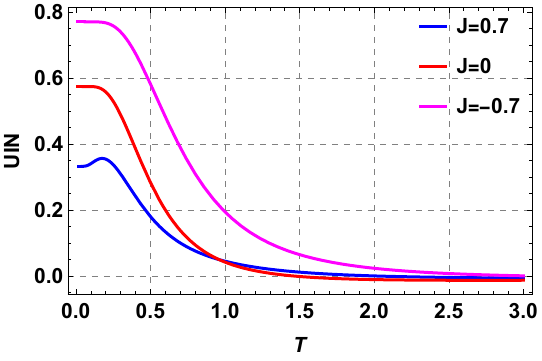}
    \caption{UIN}
    \label{Fig1c}
\end{subfigure}
\hfill
\begin{subfigure}[b]{0.45\textwidth}
    \centering
    \includegraphics[width=\textwidth, height=130px]{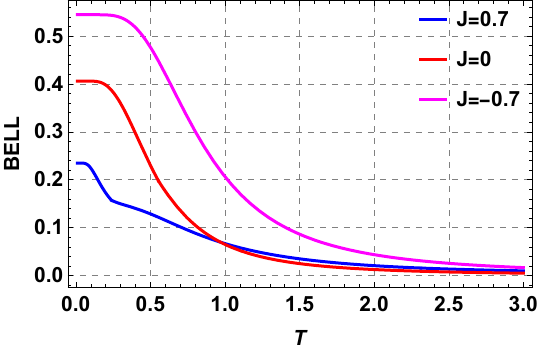}
    \caption{Bell (CHSH)}
    \label{Fig1d}
\end{subfigure}
\caption{Thermal quantum correlations for $B_1 = 0.3, B_2 = -0.7, J_z = 1, K = 0.2, K_1 = -0.1, K_2 = 0.22, Dz = 0.32, \Gamma = -0.87, \Lambda = 0.31.$ Panels show: (a) Negativity, (b) Measurement-Induced Nonlocality (MIN), (c) Uncertainty-Induced Nonlocality (UIN), and (d) Bell (CHSH) parameter.}
\label{Fig2}
\end{figure}

Further, in Fig.~2 (a–d), we study the effect of increasing temperature while varying the isotropic exchange coupling $J$ from the antiferromagnetic to the ferromagnetic regime, while keeping $J_z=1$ and all other parameters fixed.  For $J=0$, the spins are uncoupled beyond the $s_zS_z$ term. As $J$ becomes increasingly negative (ferromagnetic $XY$ coupling), negativity increases, reflecting the stronger singlet-like correlations favored by the transverse exchange. In this case as well, entanglement death is observed with rising temperature, indicating that Negativity is highly susceptible to thermal fluctuations.  MIN follows a qualitatively similar trend but remains nonzero even for smaller values of $J$ at higher $T$, underscoring its sensitivity to nonclassical disturbances beyond strict entanglement. The high $T$ “tail” of MIN is notably broader. At $T=0$, UIN again starts from a higher baseline and shows no abrupt thresholds. Its monotonic behavior across the entire $T$-range demonstrates its robustness as a measure of quantum correlations derived from information-theoretic uncertainty. Even at low temperature, a minimum ferromagnetic strength of $J=-0.7$ is not sufficient for the CHSH parameter to exceed the classical limit. With the variation of $J$, the fragility of the quantumness measures follows the same hierarchy as in the previous case, namely $\text{Bell nonlocality} \subseteq \text{Negativity} \subseteq \text{UIN} (\text{MIN})$. This demonstrates that Bell nonlocality is the most challenging to sustain, while MIN and UIN serve as more reliable indicators of nonclassical correlations even at finite temperatures.

\begin{figure}[!h]
	\centering
\begin{subfigure}[b]{0.45\textwidth}
    \centering
    \includegraphics[width=\textwidth, height=130px]{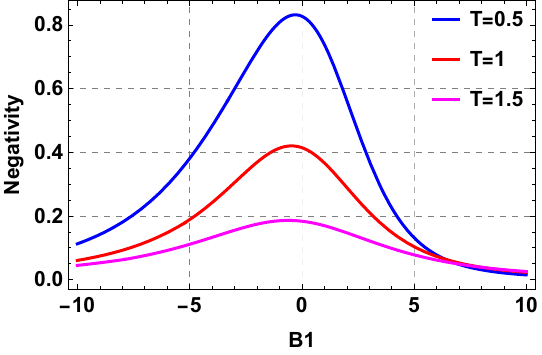}
    \caption{Negativity}
    \label{Fig1a}
\end{subfigure}
\hfill
\begin{subfigure}[b]{0.45\textwidth}
    \centering
    \includegraphics[width=\textwidth, height=130px]{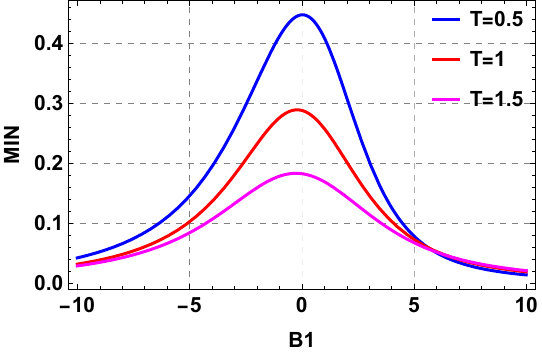}
    \caption{MIN}
    \label{Fig1b}
\end{subfigure}

\vspace{0.3cm}

\begin{subfigure}[b]{0.45\textwidth}
    \centering
    \includegraphics[width=\textwidth, height=130px]{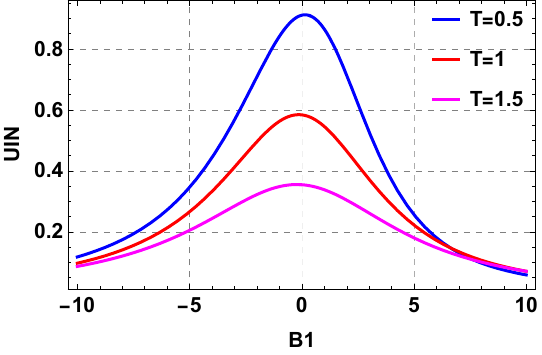}
    \caption{UIN}
    \label{Fig1c}
\end{subfigure}
\hfill
\begin{subfigure}[b]{0.45\textwidth}
    \centering
    \includegraphics[width=\textwidth, height=130px]{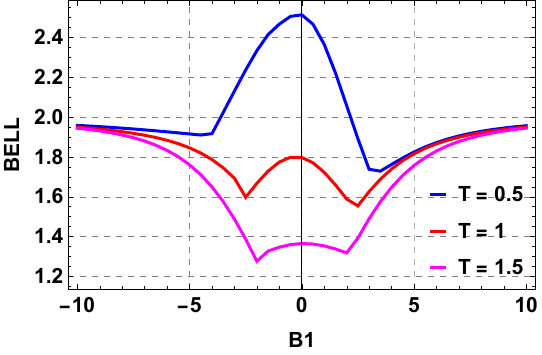}
    \caption{Bell (CHSH)}
    \label{Fig1d}
\end{subfigure}
\caption{Variation of Negativity, MIN, UIN, and Bell nonlocality (CHSH parameter) as a function of the external magnetic field $B_1$. The fixed parameters are 
 $B_2 = 0, J = -2.5, J_z = -1, K = 0.2, K_1 = -0.1, K_2 = 0.22, Dz = 0.32, \Gamma = -0.87, \Lambda = 0.31.$ Panels show: (a) Negativity, (b) Measurement-Induced Nonlocality (MIN), (c) Uncertainty-Induced Nonlocality (UIN), and (d) Bell (CHSH) parameter.}
\label{Fig3}
\end{figure}%

In Fig.~3, we plot the different quantifiers as functions of the local magnetic field $B_1$  for various temperatures ($T=0.5,1,1.5$), with the qutrit field  $B_2=0$. It is observed that the local field controls the low-temperature entanglement produced by the strong ferromagnetic transverse coupling $J=-2.5$ and negative $J_z$. Negativity attains its maximum value at the lowest temperature and decreases as the temperature increases. Furthermore, the entanglement decreases continuously with increasing $B_1$ and at sufficiently high values of the qubit magnetic field, the qubit–qutrit entanglement vanishes entirely. In Fig.~3(b), the variation of MIN with the local field $B_1$ is shown. MIN remains nonzero over a broader $B_1$ range compared to Negativity and decays more gently with lower temperature, again indicating the survival of nonclassical correlations beyond entanglement. UIN begins at the highest baseline among the three measures and exhibits the most thermal robustness, remaining appreciable even when both Negativity and MIN are suppressed. Bell nonlocality is the most fragile: violation of the CHSH bound ($>2$) is observed only at low temperature ($T=0.5$) and within a relatively narrow $B_1$. {\color{black} For the same parameters, Ref.~\cite{Yurischev2025} finds abrupt changes in the LQU and LQFI as the qubit field is varied, appearing symmetrically about $B_1=0$ in their plots. By contrast, our Negativity, MIN, and UIN vary smoothly and do not exhibit such behaviour. The CHSH parameter shows an abrupt change at comparable field magnitudes (also symmetric about $B_1=0$), but this feature does not consistently correspond to a stable Bell violation.
}

\begin{figure}[!h]
\centering
\begin{subfigure}[b]{0.45\textwidth}
    \centering
    \includegraphics[width=\textwidth, height=130px]{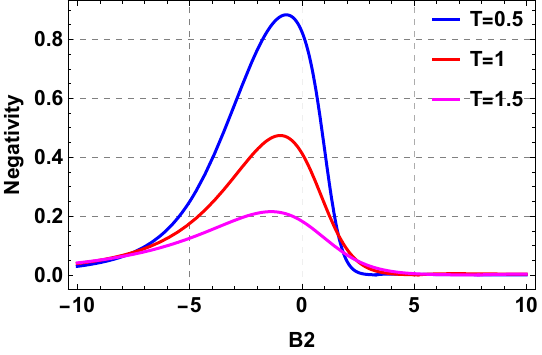}
    \caption{Negativity}
    \label{Fig1a}
\end{subfigure}
\hfill
\begin{subfigure}[b]{0.45\textwidth}
    \centering
    \includegraphics[width=\textwidth, height=130px]{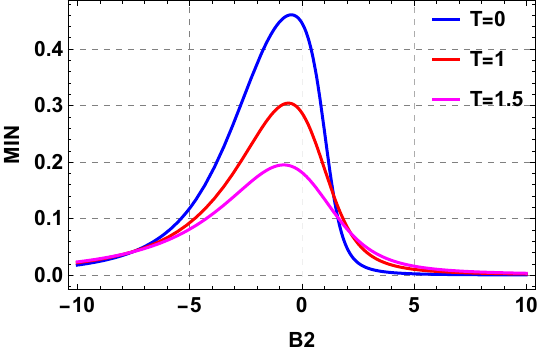}
    \caption{MIN}
    \label{Fig1b}
\end{subfigure}

\vspace{0.3cm}

\begin{subfigure}[b]{0.45\textwidth}
    \centering
    \includegraphics[width=\textwidth, height=130px]{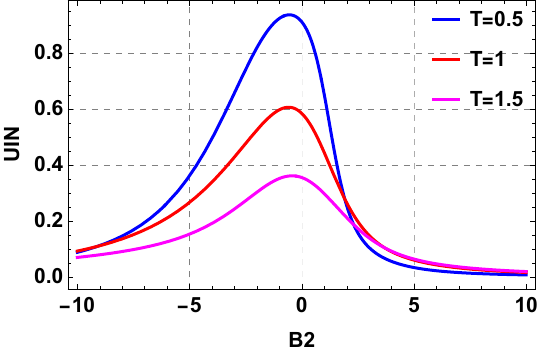}
    \caption{UIN}
    \label{Fig1c}
\end{subfigure}
\hfill
\begin{subfigure}[b]{0.45\textwidth}
    \centering
    \includegraphics[width=\textwidth, height=130px]{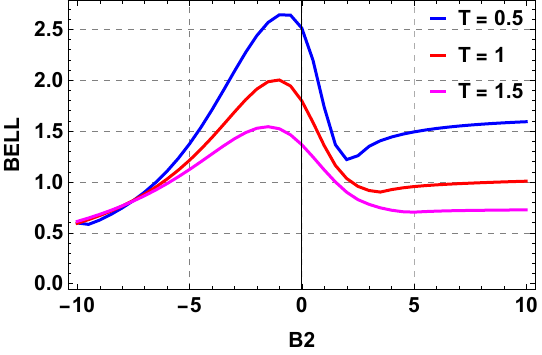}
    \caption{Bell (CHSH)}
    \label{Fig1d}
\end{subfigure}
\caption{Thermal quantum correlations for $B_1 = 0, J = -2.5, J_z = -1, K = 0.2, K_1 = -0.1, K_2 = 0.22, Dz = 0.32, \Gamma = -0.87, \Lambda = 0.31.$ Panels show: (a) Negativity, (b) Measurement-Induced Nonlocality (MIN), (c) Uncertainty-Induced Nonlocality (UIN), and (d) Bell (CHSH) parameter.}
\label{Fig4}
\end{figure}

Next, we investigate the effects of the qutrit magnetic field $B_2$ while setting the qubit field $B_1=0$ in Fig.~\ref{Fig4}. We plot all four measures as functions of $B_2$ for fixed temperatures $T=0.5, 1$ and $T=1.5$. At low temperatures, entanglement is weaker and confined to a narrower $B_2$ interval compared with the $B_1$ (Fig.~\ref{Fig3}). The Negativity is rapidly suppressed by thermal mixing and becomes nearly negligible at higher temperatures. In contrast, the MIN persists over a broader $B_2$ range than entanglement, retaining nonzero values even when the Negativity is low for higher temperatures. UIN shows the slowest thermal decay and the largest $B_1$ baseline, reflecting its information-theoretic robustness. The CHSH parameter violates the classical bound only at the lowest temperature and only within a narrow $B_2$ window; compared with the $ B_1$ scan, the observable region of Bell nonlocality is smaller. We observe that the resources attain their maximum values when the field $B_2=0$, and these maximal values decrease as the temperature increases. {\color{black} Ref.~\cite{Yurischev2025} finds an asymmetric behaviour about $B_2=0$ in the LQU and LQFI, with only a single abrupt transition when the field $B_2$ is applied. Our results display a similar qualitative structure, where we also observe an
asymmetric response and a single threshold feature, where the correlations show a sharp drop near a comparable critical value of $B_2$.}

  Comparing Figs.~\ref{Fig3} and \ref{Fig4} shows that for the strong ferromagnetic transverse coupling ($J=-2.5$)
and negative longitudinal exchange ($J_z=-1$), tuning the qubit Zeeman term ($B_1$) produces larger low-temperature
entanglement and a wider Bell-violation window than tuning the qutrit Zeeman term ($B_2$). In both scans, MIN and UIN
remain appreciable at higher temperatures, where entanglement and Bell nonlocality have already been suppressed, with UIN being the most thermally robust information-theoretic signature. The qualitative fragility ordering Bell $\subseteq$ Negativity $\subseteq$ UIN (MIN) remains valid across both
parameter sweeps. However, the exact field windows and temperature thresholds differ depending on whether the qubit
or qutrit field is being varied.

\begin{figure}[!h]
\centering
\begin{subfigure}[b]{0.45\textwidth}
    \centering
    \includegraphics[width=\textwidth, height=130px]{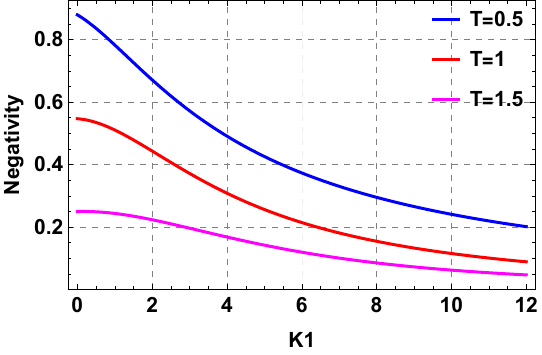}
    \caption{Negativity}
    \label{Fig1a}
\end{subfigure}
\hfill
\begin{subfigure}[b]{0.45\textwidth}
    \centering
    \includegraphics[width=\textwidth, height=130px]{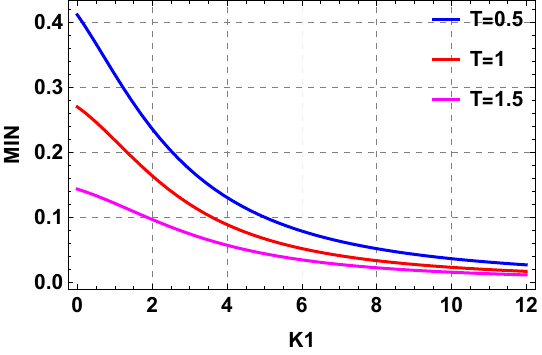}
    \caption{MIN}
    \label{Fig1b}
\end{subfigure}

\vspace{0.3cm}

\begin{subfigure}[b]{0.45\textwidth}
    \centering
    \includegraphics[width=\textwidth, height=130px]{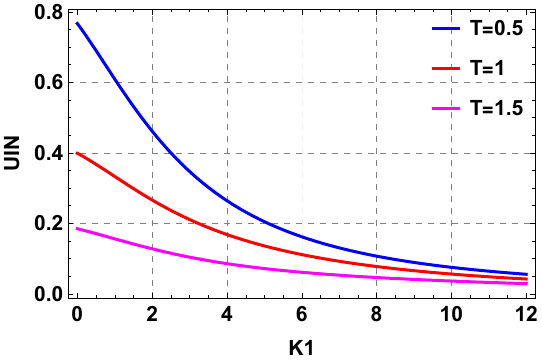}
    \caption{UIN}
    \label{Fig1c}
\end{subfigure}
\hfill
\begin{subfigure}[b]{0.45\textwidth}
    \centering
    \includegraphics[width=\textwidth, height=130px]{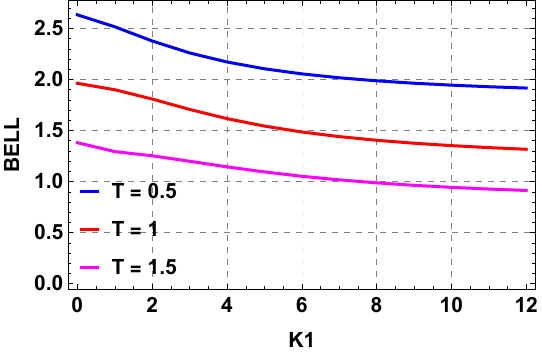}
    \caption{Bell (CHSH)}
    \label{Fig1d}
\end{subfigure}
\caption{Thermal quantum correlations for $B_1 = 0.3, B_2 = -0.7, J = -1.4, J_z = 1, K = 0.2, K_2 = 0.22, Dz = 0.32, \Gamma = -0.87, \Lambda = 0.31.$ Panels show: (a) Negativity, (b) Measurement-Induced Nonlocality (MIN), (c) Uncertainty-Induced Nonlocality (UIN), and (d) Bell (CHSH) parameter.}
\label{Fig5}
\end{figure}

Figure 5 illustrates the effect of varying the planar single-ion anisotropy parameter $K_1$ on different quantum correlation measures. Temperature primarily suppresses the entanglement/Bell windows, while MIN and UIN remain the most resilient.  Negativity, MIN, and UIN exhibit similar dependence on $K_1$: the lowest-temperature curve ($T=0.5$) shows the highest correlations, which decrease as $T$ increases, leading to correlation death at progressively smaller $K_1$ for high $T$. Bell nonlocality is the most fragile where only the lowest-temperature curve attains values above the classical bound, and then only within a narrow $K_1$ window.

\begin{figure}[!h]
\centering
\begin{subfigure}[b]{0.45\textwidth}
    \centering
    \includegraphics[width=\textwidth, height=130px]{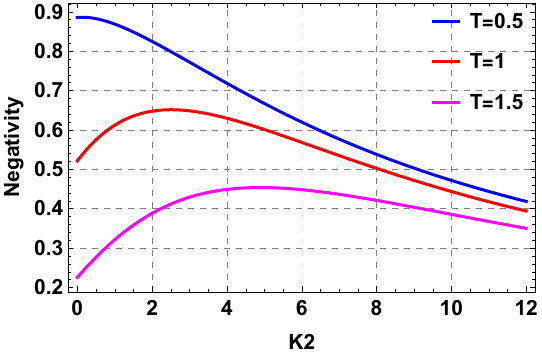}
    \caption{Negativity}
    \label{Fig1a}
\end{subfigure}
\hfill
\begin{subfigure}[b]{0.45\textwidth}
    \centering
    \includegraphics[width=\textwidth, height=130px]{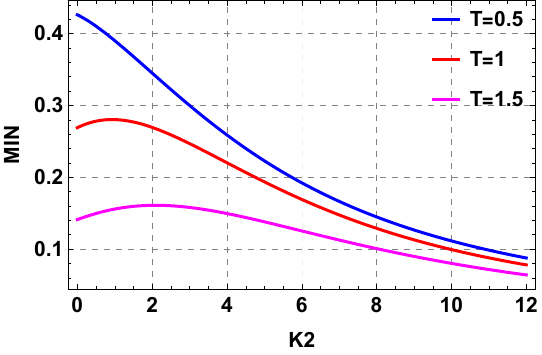}
    \caption{MIN}
    \label{Fig1b}
\end{subfigure}

\vspace{0.3cm}

\begin{subfigure}[b]{0.45\textwidth}
    \centering
    \includegraphics[width=\textwidth, height=130px]{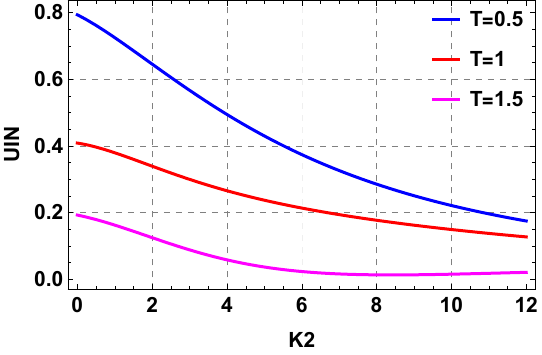}
    \caption{UIN}
    \label{Fig1c}
\end{subfigure}
\hfill
\begin{subfigure}[b]{0.45\textwidth}
    \centering
    \includegraphics[width=\textwidth, height=130px]{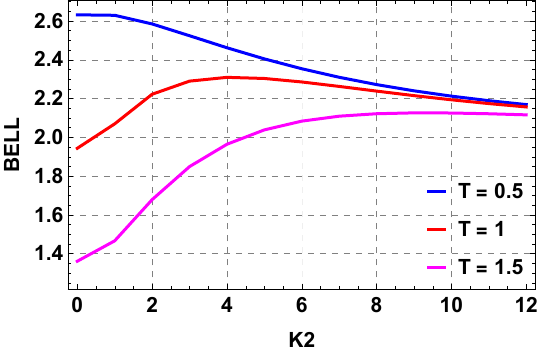}
    \caption{Bell (CHSH)}
    \label{Fig1d}
\end{subfigure}
\caption{Thermal quantum correlations for $B_1 = 0.3, B_2 = -0.7, J = -1.4, J_z = 1, K = 0.2, K_1 = -0.1, Dz = 0.32, \Gamma = -0.87, \Lambda = 0.31.$ Panels show: (a) Negativity, (b) Measurement-Induced Nonlocality (MIN), (c) Uncertainty-Induced Nonlocality (UIN), and (d) Bell (CHSH) parameter.}
\label{Fig6}
\end{figure}

In Fig. 6, we investigate the effect of scanning the biquadratic anisotropy $K_2$, which produces a stronger enhancement of entanglement and Bell nonlocality compared to $K_1$. Bell violation is observable at low temperatures for sufficiently large $K_2$. Negativity noticeably increases and decreases with $K_2$ at high temperatures. The lowest-temperature curve exhibits the highest correlations, whereas higher temperatures generally suppress correlations. The enhancement of correlations with $K_2$ is stronger than that seen in the $K_1$-scan, indicating that the biquadratic term stabilizes singlet-like correlations in this regime. Bell (CHSH) displays a wider and stronger low-temperature violation under $K_2$-variation than under $K_1$-variation. At the lowest temperature, the CHSH value exceeds the classical bound across a broader $K_2$ interval, while for higher $T$ the violation window narrows but is notably present at higher $K_2$. {\color{black} Ref. \cite{Yurischev2025} finds that both LQU and LQFI undergo a single sudden change at a specific value of the anisotropy $K_2$. In our results, we similarly observe a pronounced transition in Negativity and Bell–CHSH at a certain $K_2$, indicating a similar sensitivity to this anisotropy.}
\section{Conclusions and outlook}
\label{conclusions}

In this work, we analyzed the behavior of different quantum correlation measures in a qubit–qutrit spin system incorporating axial and planar single-ion anisotropies and external magnetic fields. By systematically studying the effects of anisotropy parameters and temperature, we identified clear distinctions in the robustness of different quantifiers. Our findings reveal that entanglement, although a fundamental resource, is highly fragile: thermal noise and anisotropy drive it to sudden death at moderate parameter values. Bell nonlocality is even more vulnerable, disappearing entirely except in very narrow low-temperature regimes. 

\textcolor{black}{The present investigation, together with a comparative analysis with the LQU/LQFI results of Yurischev et al. \cite{Yurischev2025}, reveals clear differences in robustness of discord-like measures against entanglement and Bell nonlocality. Specifically, discord-like measures MIN and UIN decay much more slowly with temperature: even after entanglement vanishes (Negativity = 0), both MIN and UIN remain appreciable, whereas the Bell–CHSH value falls below the classical bound of 2 and is lost at relatively low thermal noise levels. We observe the fragility hierarchy,
\[
\text{Bell nonlocality} \subseteq \text{Negativity} \subseteq \text{UIN} (\text{MIN}),
\]
which provides a unified picture of correlation strength in anisotropic hybrid spin systems. Importantly, this ordering aligns with the general correlation hierarchy under decoherence known from prior works. This fragility ordering has been noted before by Ali et al., who discuss the general view of ``$\text{Bell nonlocality} \subseteq \text{Quantum steering} \subseteq \text{Entanglement} \subseteq \text{Discord} \subseteq \text{Coherence}$" in a spin-1/2 model under thermal regime \cite{Ali2024}. Likewise, Paulson et al. have shown that under both Markovian and non-Markovian noise, the decay and revival of correlations follow this hierarchy \cite{Paulson2021}. Similarly, Atta et al. found that under classical dephasing channels, an “average steered coherence” (a coherence-type quantifier) survives longer than Bell nonlocality or steerability \cite{rahman2023}. We previously investigated correlated amplitude-damping channels using weak-measurement reversal protocols and reported the same nested fragility ordering, with discord/coherence surviving longer than entanglement, thereby extending the observed hierarchy to correlated noise scenarios \cite{Abhignan2025}. Beyond these two-qubit correlation comparisons, the present results for the axially symmetric qubit–qutrit system confirm that the pattern is consistent within the general view of the fragility hierarchy.}

\textcolor{black}{These insights emphasize that while entanglement remains a central concept, other nonclassical quantifiers such as MIN and UIN may play a more crucial role in realistic qubit–qutrit implementations, where environmental decoherence and anisotropy cannot be neglected. Although entanglement and Bell nonlocality have been studied in Nuclear Magnetic Resonance (NMR) systems~\cite{Silva_2013}, solid-state spins~\cite{PMID:26503041}, and superconducting qubits~\cite{BNLsupcondcircs}, no experiments have yet analyzed these discord-like measures. Future studies may extend this quantum discord framework to larger spin systems \cite{Lesanovsky_2019}, non-Markovian environments \cite{Benabdallah2025adp}, and experimentally relevant platforms such as NMR spin ensembles \cite{Xu_2012}, molecular magnets \cite{Naveena2022}, and ultracold Rydberg arrays \cite{Lesanovsky_2019}, thereby bridging theoretical predictions with realizable quantum technologies.}


\section*{Data availability} 

\noindent All data generated or analysed during this study are included in this published article.

\bibliographystyle{ieeetr}
\bibliography{Reference}

\end{document}